\documentstyle[aps]{revtex}

\begin{document}


\draft

\title{Angular momentum near the black hole threshold in scalar field
collapse}

\author{David Garfinkle}
\address{Dept. of Physics, Oakland University, Rochester, MI 48309,
USA}

\author{Carsten Gundlach}
\address{Max-Planck-Institut f\"ur Gravitationsphysik
(Albert-Einstein-Institut), Schlaatzweg 1, 14473 Potsdam, Germany}

\author{Jos\'e M. Mart\'\i n-Garc\'\i a}
\address{Laboratorio de Astrof\'\i sica Espacial y F\'\i sica Fundamental,
Apartado 50727, 28080 Madrid, Spain}

\date{2 December 1998}

\maketitle


\begin{abstract}

For the formation of a black hole in the gravitational collapse of a
massless scalar field, we calculate a critical exponent that governs
the black hole angular momentum for slightly non-spherical initial
data near the black hole threshold. We calculate the scaling law by
second-order perturbation theory. We then use the numerical results of
a previous first-order perturbative analysis to obtain the numerical
value $\mu\simeq0.76$ for the angular momentum critical exponent. A
quasi-periodic fine structure is superimposed on the overall power
law.

\end{abstract}


\section{Introduction}


In the space of initial data for a self-gravitating system, the black
hole threshold is the limit between initial data that eventually form
a black hole and initial data that do not. More precisely, if one
considers smooth one-parameter families of initial data such that for
large values of the parameter $p$ a black hole is formed, but not for
small values, then there is a critical value $p_*$ (that can be found
by bisection) where a black hole is first formed. Such data are called
critical data. All numerical experiments are compatible with the
picture that the critical data form a smooth hypersurface of
codimension one, called the critical surface, in the
infinite-dimensional space of initial data.

The evolution of any data sufficiently close to the black hole
threshold shows what is now commonly known as critical phenomena in
gravitational collapse.  The solution evolved from near-critical data,
with $p\simeq p_*$, approaches a universal intermediate attractor in a
region of spacetime. This universal ``critical solution'' has a
discrete or continuous self-similarity. Numerical experiments as well
as perturbative calculations are compatible with the picture that it
is an attractor inside the critical surface, with precisely one
unstable perturbation mode pointing out of the critical surface.

If $p>p_*$, and a black hole is formed, its mass scales with $p$ as
\cite{Choptuik93}
\begin{equation}
\label{mass}
M\simeq C\, (p-p_*)^\gamma,
\end{equation}
where $\gamma$ is the same for all families of initial data.  However,
scaling is not only a supercritical property. Any dimensionful
variable of the system scales with $p-p_*$. For example the maximum
curvature of spacetime scales as $(p_*-p)^{-2\gamma}$ for $p<p_*$
\cite{Garfinkle}. The critical exponent $\gamma$ is directly related
to the growth rate of the one unstable mode.

The critical exponent $\gamma$ is independent of the initial data. It
depends on the type of matter (or absence of matter, in the collapse
of pure gravitational waves), but is independent of any dimensionful
constants in the matter equations of motion. Critical phenomena were
first observed in the spherically symmetric collapse of a scalar field
\cite{Choptuik93} and the axisymmetric collapse of gravitational waves
\cite{AbrahamsEvans}. The structure of the intermediate attractor that
plays the role of the critical point was clarified in
\cite{EvansColeman} and \cite{Gundlach_Chop1}.  The critical exponent
for the black hole mass was calculated by perturbation theory around
the critical point in \cite{KoikeHaraAdachi}, and more generally in
\cite{Gundlach_Chop2}. An independent critical exponent for the black
hole electric charge was predicted in \cite{GundlachMartin} and
subsequently measured in \cite{HodPiran_charge}. The prediction of a
critical exponent for the black hole's angular momentum required
perturbation theory beyond spherical symmetry, and was first carried
out in \cite{Gundlach_nonspherical,Gundlach_angmom} for perfect fluid
collapse. A detailed review of all this can be found in
\cite{Gundlach_critreview}.

Now two of us have calculated the non-spherical linear perturbations
around the critical point in scalar field collapse
\cite{critscalar}. In the present paper, we build on that work to
quantitatively predict the scaling of the black hole angular momentum
in critical scalar field collapse. The perfect fluid and scalar field
cases have in common that the critical spacetime is spherically
symmetric. Black hole angular momentum, which breaks spherical
symmetry, can therefore be treated perturbatively.  There are two
important differences, however.

The first difference is this. The critical point spacetime in perfect
fluid collapse has, apart from spherical symmetry, another continuous
symmetry, namely continuous self-similarity (CSS), also called a
homothety. Roughly speaking, this means that it does not depend on $t$
and $r$ separately, but only through the dimensionless combination
$x=-r/t$ (for a suitable choice of coordinates that we do not need to
discuss here). The perturbations of this solution must then have the
form ${\rm Re}\left[(-t)^{-\lambda} f(x)\right]$, where both the
eigenvalue $\lambda$ and the mode function $f(x)$ can be complex. In
the perfect fluid case the $\lambda$ of the perturbation that
dominates angular momentum -- the $l=1$ axial mode with the largest
real part -- is real. (The statement in \cite{Gundlach_angmom} that
$\lambda$ is complex was wrong.) The black hole angular momentum in
critical perfect fluid collapse scales as
\begin{equation}
\label{CSS_L}
\vec L \simeq \vec L_0
(p-p_*)^\mu , \quad \text{(perfect fluid, CSS)}
\end{equation}
In contrast, the scalar field critical spacetime depends not only on
$x$, but in a restricted way also on $t$: it is periodic in $\tau = -
\ln(-t)$. (Again this holds only in suitable coordinates). This
symmetry is called discrete self-similarity (DSS)
\cite{Gundlach_Chop2}. Furthermore the critical exponent itself is
complex.  The imaginary part of the critical exponent and the
independent period $\Delta$ in $\tau$ of the background critical point
solution then combine to give an overall power law modified by
quasiperiodic behavior, as we shall see below.

The second new difference between the fluid and the scalar field is
that angular momentum can be treated in first-order perturbation
theory around spherical symmetry for perfect fluid collapse, but not
for scalar field collapse. We have to go to second-order perturbation
theory to obtain a non-zero effect. Fortunately the relevant
second-order perturbation degree of freedom does not obey a wave-like
equation which would require independent free initial data, but is
totally determined by its source, which is quadratic in first-order
perturbations. This allows us to obtain the desired result with a
calculation based on dimensional analysis and selection rules for the
angular dependency of the perturbations. At the end of the calculation
we insert the numerical value for a particular complex eigenvalue
$\lambda$ into our result to obtain a numerical value for the angular
momentum critical exponent.

The paper is structured as follows. First we define perturbation
theory around spherical symmetry to all orders and show, both formally
and intuitively, why we have to go beyond first-order
perturbations. Then we consider the second-order perturbations that
give rise to angular momentum. In order to make our presentation more
self-contained, we briefly review how the mass scaling law is
derived. Then we derive the angular momentum scaling law for a
spherical, DSS, background, and insert numbers for the scalar field
case.


\section{First-order perturbations around spherical symmetry}


In order to define higher order perturbation theory, we formally
expand the metric and the scalar field as
\begin{eqnarray}
g_{\mu\nu}(\epsilon) & = & g_{\mu\nu}^{(0)} + \epsilon g_{\mu\nu}^{(1)}
+ \epsilon^2 g_{\mu\nu}^{(2)} + O(\epsilon^3), \\
\phi(\epsilon) & = & \phi^{(0)} + \epsilon \phi^{(1)}
+ \epsilon^2 \phi^{(2)} + O(\epsilon^3).
\end{eqnarray}
In the following we use the shorthand $u$ to denote both $g_{\mu\nu}$
and $\phi$. If we write the field equations, that is, the Einstein
equations and the scalar wave equation, formally as
\begin{equation}
{\cal E}(u)=0, \qquad u=(g_{\mu\nu},\phi)=u(\epsilon), 
\end{equation}
and take formal derivatives with respect to $\epsilon$, we obtain as
the leading orders
\begin{eqnarray}
&& {\cal E}(u^{(0)}) = 0, \\
\label{pert1}
&& {\cal L}(u^{(1)}) = 0, \\
\label{pert2}
&& {\cal L}(u^{(2)}) = {\cal S}^{(2)}(u^{(1)},u^{(1)}),
\end{eqnarray}
where $\cal L$ is a linear derivative operator, and $S$ is a quadratic
derivative operator. 

We perturb around a spherically symmetric solution $u^{(0)}$. In this
paper we consider a double perturbation expansion, both around
spherical symmetry, and around criticality. The small parameter
$\epsilon$ in the following always refers to deviations from spherical
symmetry. (The small deviation from criticality will be parameterized
by $p-p_*$.) 

We begin by establishing that no black hole angular momentum can arise
in first-order perturbation theory, for a spherically symmetric
background solution with only scalar field matter. We begin by noting
\cite{Gleiser_etal} that the Kerr metric with mass $M$ and angular
momentum $L$ in Boyer-Lindquist coordinates can be written as a
perturbation of the Schwarzschild metric (for simplicity we assume
that the angular momentum is in the $z$-direction):
\begin{eqnarray}
g^{(0)}_{\mu\nu} dx^\mu \, dx^\nu &=&-\left(1-{2M\over r}\right) dt^2
+ \left(1-{2M\over r}\right)^{-1} dr^2 + r^2 \left(d\theta^2 + \sin^2\theta
\,d\varphi^2\right), \\
\label{g1}
g^{(1)}_{\mu\nu} dx^\mu \, dx^\nu &=&- 4{L\over r} \sin^2 \theta \,
dt\, d\varphi = 4{L\over r} \sin \theta {\partial \over \partial
\theta} \sqrt{4\pi\over 3} Y_{10} (\theta,\varphi) \,dt\, d\varphi ,
\end{eqnarray}
where $Y_{10}$ is the spherical harmonic with $l=1$ and $m=0$.  Note
that this is the only first-order metric perturbation when we expand
in powers of $L$. The other metric coefficients are changed only from
$O(L^2)$ on. 

We adapt the gauge-invariant perturbation framework of Gerlach and
Sengupta \cite{GS1}, which is reviewed in more detail in
\cite{critscalar}.  In this framework, all four-dimensional tensor
perturbations are covariantly decomposed into series of tensorial
spherical harmonics, which carry all the angular dependence, with
coefficients which depend on $r$ and $t$. These harmonics are tensors
derived from the $Y_{lm}$ and their derivatives, living on the
two-spheres of spherical symmetry (coordinates $\theta$ and $\varphi$,
tensor indices $a$). The coefficients of those expansions are tensors
living in the two-dimensional reduced manifold (coordinates $r$ and
$t$, tensor indices $A$). This decomposition is performed because
around spherical symmetry different $l,m$ components of the expansion
decouple, so we can study each case separately. Each $l,m$ linear
perturbation decouple further into two parts: axial [with parity
$(-1)^{l+1}$] and polar [with parity $(-1)^l$].  

In this notation, the linear metric perturbation (\ref{g1}) is written
as
\begin{equation}
g^{(1)}_{\mu\nu} dx^\mu \, dx^\nu = 2 k_A^{(1)} dx^A \, S_{10a} dx^a,
\end{equation}
where $S_{10a}$ is the axial harmonic vector field with angular
dependence $l=1$, $m=0$, formed by the general rule
$S_{lma}={\epsilon_a}^bY_{lm:b}$. (Here $:a$ is the covariant
derivative on the unit two-sphere, and $\epsilon_{ab}$ is the
corresponding unit antisymmetric tensor.) The angular momentum
perturbation $L_z$ we are interested in can therefore be characterized
as the axial $l=1$, $m=0$, gauge-invariant metric perturbation. ($m=1$
and $m=-1$ parameterize $L_x$ and $L_y$.) By comparison with
(\ref{g1}) we read off that
\begin{equation}
k_A^{(1)} dx^A = \sqrt{4\pi\over 3} {2L\over r} dt
\end{equation}
for $l=1$ and $m=0$.  $k^{(1)}_A$ is only partially gauge-invariant
for $l=1$, but its curl $\Pi^{(1)}$ is gauge-invariant for all $l>0$,
and contains all the gauge-invariant information. We calculate
\begin{equation}
\label{KerrPi}
\Pi^{(1)} = \epsilon^{AB}(r^{-2}k_A^{(1)})_{|B} = - \sqrt{4\pi\over 3} {6L\over r^4},
\end{equation}
for $l=1$ and $m=0$. (In the notation of \cite{critscalar} these
objects are simply called $k_A$ and $\Pi$ because there we only work
with first-order perturbations.) $|A$ is the covariant derivative on
the reduced manifold, and $\epsilon^{AB}$ the corresponding unit
antisymmetric tensor.  

Let us compare this particular solution with the general linearized
equation (\ref{pert1}) it must obey. For $l\ge 2$, $\Pi^{(1)}$ obeys a
wave equation, but for $l=1$ it obeys a first-order differential
equation that can be integrated trivially to yield
\begin{equation}
\label{EEl=1}
r^4 \Pi^{(1)} - 16\pi T^{(1)} - c =0,
\end{equation}
where $T^{(1)}$ is a scalar constructed from the axial gauge invariant
matter perturbations, and $c$ is an integration constant. Note that
$\cal L$ degenerates from a derivative operator to an algebraic
operator for $l=1$ axial perturbations. In spite of its trivial
appearance, equation (\ref{EEl=1}) is the linearized Einstein equation
relating gauge-invariant axial $l=1$ metric perturbations to their
matter sources.

If the spacetime is vacuum, the spherical background must be
Schwarzschild, and the matter perturbation $T^{(1)}$ vanishes. The only
axial physical perturbation is then the one that takes
Schwarzschild infinitesimally into Kerr. We see that for this
perturbation
\begin{equation}
T^{(1)}=0, \qquad c = -  6 \sqrt{4\pi\over 3} L.
\end{equation}
(There is one other physical perturbation of Schwarzschild, $l=0$
polar, which changes the mass of the black hole.)

If we demand regularity at the center $r=0$, as we do here, we must
have $c=0$, and this is also true in the presence of matter. The
scalar field and its perturbation (to all orders) are polar, simply
because they are spacetime scalars. Therefore $T^{(1)}$ vanishes for
scalar field matter and we have $\Pi^{(1)}=0$. (To the next order,
polar and axial perturbations do mix, so that to second order $T$ and
hence $\Pi$ do not vanish. This will be discussed in the next
section.)

While this is the complete argument for the absence of black hole
angular momentum in the first-order perturbation calculation, we have
also found two intuitive ones. Angular momentum is present in the
first-order perturbations of spherical fluid collapse, but not of
scalar field collapse, because the fluid is made up of individual
particles with a rest mass which can go round while the entire
configuration remains axisymmetric. The angular momentum is
proportional to the tangential velocity component $u^\varphi$. The
scalar field has no such particles, and we need to make spoke-like
structures in the field (thus breaking axisymmetry) and then make
these go round. Intuitively, we need two powers of $\epsilon$ to do
this: one to make spokes, and another to make them go round. This
argument is backed up by the observation that the conserved angular
momentum (Noether charge) of a scalar field on a Minkowski background,
for example the $z$-component, is simply
\begin{equation}
L_{\text{$\phi$, Minkowski}}^z = \int \phi_{,t} \phi_{,\varphi} \,
d^3x = O(\epsilon^2).
\end{equation}
One can easily see that this is again quadratic in deviations from
spherical symmetry.

A second intuitive argument for the absence of first-order angular
momentum perturbations in the scalar field comes precisely from its
partial equivalence with a perfect fluid (with $p=\rho$): the
4-velocity of that pseudo-fluid is the normalized gradient of the
scalar field, and therefore irrotational.


\section{Second-order perturbations}


In second-order perturbation theory, we can define second-order
versions of the gauge-invariant perturbations. (Equivalently, we could
fix the gauge separately at each order.) Independent of the detailed
field equations, the term in the $n$-th order equations that is linear
in the $n$-th order perturbations is the same at all orders, namely
$\cal L$. For $l=1$ axial perturbations, equation (\ref{pert2})
therefore takes the simple algebraic form
\begin{equation}
r^4 \Pi^{(2)} - 16\pi T^{(2)} - c = S^{(2)}\left(u^{(1)},u^{(1)}\right)
\end{equation}
where $\Pi^{(2)}$ and $T^{(2)}$ are just $\Pi^{(1)}$ and $T^{(1)}$
with $u^{(1)}$ replacing $u^{(2)}$. Just as for the first-order
perturbations, $T^{(2)}$ vanishes identically for scalar field
matter, and the integration constant $c$ vanishes if we consider only
solutions with a regular center $r=0$.

But now there is also the source $S^{(2)}$, which is quadratic in
first order perturbations $u^{(1)}$. It does not vanish in general,
and generates a non-vanishing $\Pi^{(2)}$. Therefore it can introduce
angular momentum. If and when a black hole is formed, it must settle
down to the Kerr solution at late times. $S^{(2)}$ must then approach
a constant at late times outside the horizon, thus transforming a
Schwarzschild into a Kerr black hole. Inside the horizon, where the
singularity forms, perturbation theory around a regular center will
break down, but that does not affect our calculations.

Finding the expression for $S^{(2)}$ would require more effort than
writing down the left-hand side, but fortunately we are interested
only in scaling arguments, not in the detailed behavior of $\Pi^{(2)}$
as a function of $r$ and $t$. We note that $\Pi^{(2)}$ belongs to the
axial sector, and that we are only interested in its $l=1$, $m=0$
component, as that is the one connected to black hole angular
momentum. All other second-order perturbations (except a mass
perturbation, as mentioned above) must eventually be radiated away as
the black hole settles down to the Kerr solution. We can therefore
restrict attention to terms which are quadratic in first-order
perturbations, axial, and have angular dependence $l=1$, $m=0$. The
restriction to $m=0$ is equivalent to the restriction to the
$z$-component, $L_z$, of angular momentum. We make it only for
simplicity of presentation. $L_x$ and $L_y$ are related to complex
linear combinations of $m=1$ and $m=-1$.

Let us denote the two factors $u^{(1)}$ by $u'$ and $u''$.  They must
have $m=m'+m''=0$. We do not require that the number $l'+l''$ is odd,
nor that of $u'$ and $u''$ one is polar and the other axial. An
example for the mixing of polar and axial perturbations to quadratic
order is
\begin{equation}
Y_{lm} \, Y_{l-m:a} = - im\sqrt{3\over
16\pi} S_{10a} + \text{other terms}.
\end{equation}
for any $l$ and $m$. Note that any $m'+m''=0$ gives rise to the axial
vector field $S_{10a}$ that characterizes angular momentum in the
$z$-direction, except $m'=m''=0$. That $m'=m''=0$ is excluded is also
clear on physical grounds, as we have argued above that an
axisymmetric scalar field configuration cannot have angular
momentum. In the simplest case $l=1$ the complete expression is
\begin{equation}
Y_{11} \, Y_{1-1:a} = - i\sqrt{3\over
16\pi} S_{10a} + {1\over 4\sqrt{5\pi}} Y_{20:a} .
\end{equation}
In words: two polar $l=1$ perturbations combine to give an axial $l=1$
perturbation, as well as the expected $l=2$ polar perturbation.


\section{Review of mass scaling in critical collapse}


We have clarified the role of second-order perturbation theory
for calculating angular momentum-like quantities in almost spherical
scalar field collapse. In the following we specialize to a particular
background solution, the so-called critical solution. The derivation
of the angular momentum scaling law is a continuation of the
previous derivations of scaling laws for the mass
\cite{EvansColeman,Gundlach_Chop2}, electric charge
\cite{GundlachMartin} and angular momentum \cite{Gundlach_angmom}. 
Therefore we do not try here to be fully self-contained, but rather
remind the reader of the general ideas
underlying these calculations. A more detailed review is contained in
\cite{Gundlach_critreview}.

Critical collapse is dominated by a single solution which has the two
crucial properties of being self-similar (CSS or DSS) and of having
precisely one growing perturbation mode. This solution is best given
in coordinates $x$ and $\tau$, which are (with some simplification)
$x=-r/t$ and $\tau=-\ln(-t)$. The important properties of these
coordinates are that both are dimensionless, that $x$ is invariant
under rescaling of space and time, and that $\tau$, being the
logarithm of a scale, changes only by an additive constant. The
critical scalar field solution, in coordinates $\tau,x,\theta,\varphi$
is then of the form
\begin{equation}
g_{\mu\nu} = l_0^2 e^{-2\tau} \bar g_{\mu\nu}(x,\tilde\tau),
\qquad \phi = \phi(x,\tilde\tau),
\end{equation}
where the tilde over the argument $\tau$ indicates that the conformal
metric $\bar g_{\mu\nu}$ and scalar field $\phi$ depend on $\tau$ only
periodically, with a period $\Delta$. Note that $\bar g_{\mu\nu}$,
$\phi$, and the coordinates $\tau,x,\theta,\varphi$ are conveniently
thought of as dimensionless, with the arbitrary constant scale $l_0$
the only dimensionful quantity. Note also that by construction, $l_0$
always comes together with $e^{-\tau}$. From this it follows that
in the self-similar spacetime
\begin{equation}
\label{scale}
\text{any masslike quantity} \sim l_0 e^{-\tau}.
\end{equation}

Now consider the one growing linear perturbation mode. As the
background is periodic in $\tau$, it must be of the form
\begin{equation}
\text{growing mode} \sim (p-p_*) e^{\lambda_0\tau} f_0(x,\tilde\tau).
\end{equation}
$\lambda_0$ is real and positive, so that the perturbation grows as we
approach smaller scales ($t\to 0_-,\, r\to 0,\, \tau\to\infty$). As by
assumption there is only one such mode, and the background is real,
$\lambda_0$ and $f_0$ must be real, but in general $\lambda_i$ and
$f_i$ exist in complex conjugate pairs. The overall amplitude of the
growing mode depends on the initial data in a complicated way, but to
leading order it must be proportional to $p-p_*$, because for $p=p_*$
the critical solution lives forever, by definition of being the
critical solution, and so the growing mode cannot be present.

Now consider the spacelike hypersurface $\tau=\tau_*$, with the value
of $\tau_*$ defined by
\begin{equation}
\text{growing mode} \sim (p-p_*) e^{\lambda_0\tau_*} \sim \text{some
fiducial amplitude}.
\end{equation}
For $\tau_*$ this gives
\begin{equation}
\label{taustar}
\lambda_0\tau_* = - \ln(p-p_*) + {\rm const.}
\end{equation}
At a later stage, we can no longer approximate the spacetime as
self-similar plus a perturbation, but the Cauchy data at $\tau=\tau_*$
are independent of the initial data -- the decaying perturbations have
all decayed, and the one growing perturbation has reached its fiducial
amplitude -- up to an overall scale, which must be given by
(\ref{scale}). Therefore we have for the black hole mass
\begin{equation}
M\sim l_0 e^{-\tau_*} \sim  (p-p_*)^{1\over \lambda_0},
\end{equation}
so that we have found the law (\ref{mass}) with $\gamma = 1/\lambda_0$.


\section{Angular momentum scaling in DSS critical collapse}


At the fiducial time $\tau=\tau_*$, we have for the angular momentum
of what will become the black hole
\begin{equation}
\label{quadsource}
L_z \sim S^{(2)}\left(u^{(1)},u^{(1)}\right)(\tau_*) \sim
l_0^2 e^{-2\tau_*} {\rm Re} \, u'(\tau_*) \ {\rm Re} \,u''(\tau_*) 
\sim l_0^2 e^{-2\tau_*} {\rm Re} [C'f'(\tau_*) e^{\lambda'\tau_*}] \
{\rm Re} [C''f''(\tau_*) e^{\lambda''\tau_*}]
\end{equation}
Here $\lambda'$ and $\lambda''$ are the complex exponents of the most
slowly decaying modes of two separate first-order perturbations which
are compatible with the angular dependence selection rules, and $f'$
and $f''$ are the corresponding complex mode functions, which are
periodic in $\tau$ with period $\Delta$. $C'$ and $C''$ are complex
constants that depend on the family of initial data. (They also depend
on $p$, but here we only take their leading order, which is
constant. We take the real part of $u'$ and $u''$ separately because the
complex notation is only a shorthand for sines and cosines.)  In order
to simplify we have made the jump from the $x$-dependent quantity
$S^{(2)}$ to the simple number $L$ in the first equality, and we have
therefore suppressed the $x$-dependence of $f'$ and $f''$.  Note that
the factor $l_0^2\,e^{-2\tau_*}$ in the second equality appears by
dimensional analysis -- $L$ has dimensions (length)${}^2$.

In \cite{critscalar} we found that the most slowly decaying
perturbation with $l\ne 0$ is the polar $l=2$ perturbation, which has
$\lambda\simeq-0.06\times(1/\Delta)+0.30\times (2\pi i/\Delta) \simeq
-0.017 + 0.55i$.  Therefore $L_z$ will be dominated by contributions
where both $u'$ and $u''$ are $l=2$ polar. This does not mean,
however, that $u'=u''$. $S^{(2)}$ will contain dominant contributions
from both $m'=-m''=1$ and $m'=-m''=2$. Furthermore $C'$ and $C''$ are
independent complex constants for each of these two cases. (We have
chosen our complex notation so that $C$ can take arbitrary complex
values, but $\lambda$ and its complex conjugate are not counted as
independent.) The function $L_z(p-p_*)$ is therefore parameterized by
four independent complex constants, namely the constants $C$ for
$l=2$, polar, with $m=-1,1,-2,2$.

Inserting the value (\ref{taustar}) for $\tau_*$, and putting back the
vector character of $\vec L$, and the periodic nature of the
background, we can summarize our result as
\begin{equation}
\label{DSS_L}
\vec L \simeq \vec L_0[\ln(p-p_*)]
(p-p_*)^\mu, \quad \text{(DSS)}
\end{equation}
where $\vec L_0[\ln(p-p_*)]$ is a quasiperiodic function that depends
on the family of initial data, and $\mu$ is a universal critical
exponent that is given by
\begin{equation}
\mu = {\rm Re} {2 - \lambda' - \lambda'' \over \lambda_0} 
= (2 - 2{\rm Re}\lambda') \gamma. 
\end{equation}
Note that $\lambda'$ has negative real part, so that we have $\mu>
2\gamma$. With $\gamma\simeq 0.374$ and ${\rm Re}\lambda'\simeq
-0.017$ we predict a critical exponent $\mu\simeq0.76$, which is
barely larger than $2\gamma$.

The Fourier spectrum of $\vec L_0[\ln(p-p_*)]$ with respect to its
formal argument contains the angular frequencies 
\begin{equation}
N{2\pi\gamma\over\Delta}, \qquad 
N{2\pi\gamma\over\Delta} \pm 2\gamma\,{\rm Im}\lambda'
\end{equation}
for integer $N$. We could be more precise by writing down the general
form of $\vec L_0[\ln(p-p_*)]$ as a sum involving eight universal (but
from the present calculation, unknown) periodic functions and 24 real
constants depending on the family of initial data, but that would not
be very helpful. Nevertheless, in numerical collapse simulations it
should be possible to spot not only the overall power law, but also a
fine structure with the fundamental angular frequency $2\pi\gamma/
\Delta\simeq 0.683$ and the offset $2\gamma{\rm Im}\lambda'\simeq
0.41$ in the Fourier transform of $\vec L_0$ with respect to
$\ln(p-p_*)$.

\section{Acknowledgments}

This work was partially supported by NSF Grant PHY-9722039 to Oakland
University and the 1994 Plan de Formaci\'on de Personal Investigador
of the Comunidad Aut\'onoma de Madrid.  DG and JMM would like to thank 
the Albert Einstein Institute for hospitality.
 


\end{document}